\documentclass[a4paper,12pt]{article}
\usepackage{epsfig}
\usepackage{amssymb}
\begin{document}
\thispagestyle{empty}

\newcommand{\etal}  {{\it{et al.}}}  
\def\Journal#1#2#3#4{{#1} {\bf #2}, #3 (#4)}
\def\PRD{Phys.\ Rev.\ D}
\def\NIMA{Nucl.\ Instrum.\ Methods A}
\def\PRL{Phys.\ Rev.\ Lett.\ }
\def\PLB{Phys.\ Lett.\ B}
\def\EPJ{Eur.\ Phys.\ J}
\def\IEEETNS{IEEE Trans.\ Nucl.\ Sci.\ }
\def\CPCD{Comput.\ Phys.\ Commun.\ }

\bigskip

{\bf
\begin{center}
 \textbf{\large {Two-particle "ridge" correlations } }
\end{center}
}


\begin{center}
{\large G.A. Kozlov  }
\end{center}


\begin{center}
\noindent
 {
 Bogolyubov Laboratory of Theoretical Physics\\
 Joint Institute for Nuclear Research,\\
 Joliot-Curie st., 6 Dubna\\
 141980 Moscow region, Russia
 }
\end{center}

\begin {abstract}
\noindent

We have examined on the qualitative level the  angular correlations of two particles produced in $pp$ collisions. The effective  model based on the dual description of Yang-Mills theory at finite temperature  where the scalar (dilaton) field provides the flux-tube solution nearly phase transition is used. This can explain the formation of the "ridge" structure in two-particle angular correlation function versus both the pseudorapidity $\Delta\eta$ and  the azimuthal $\Delta\varphi$ differences.
The shape of correlation pattern with $\Delta\varphi\simeq 0$ depends on the form of the correlation function and on in-medium distortion which affect the "ridge" behavior.

\end {abstract}




\bigskip

{\it Introduction.-}  The Large Hadron Collider (LHC) at CERN has already  provided particle physicists with new collection of data.
Recently, the CMS collaboration at the LHC has published [1] the first result on discovery
of the "ridge" pattern in two particle angular correlations in high multiplicity (with number
of particles $N > 110$) $pp$ collisions at the center-of-mass energy $\sqrt {s}$ = 7 TeV.

The effect of two-particle correlations is clear and undeniable part of high energy physics, complicating the quantum statistical description of multi-particle strong interactions at high temperatures and densities.

The observation of the "ridge" at the CMS has initiated a large number of papers (see, e.g., papers in [2] and the references therein) having the aim to explain this effect. The similar structure of the long-range near side correlations  has been  seen at RHIC  [3],  and one of the explanations was
based on the proposal of the explosion of high energy density matter. It is supposed that the observation of the "ridge is in the line of its interpretation as possible evidence for quark-gluon phase. In the related context, as mentioned in [4], the boost invariant picture of the particle production, and, therefore, long-range rapidity correlations, are the characteristic of the soft multiparticle production within the Lund string model [5].

In this paper, we examine the possibility that $pp$ collisions with high rate of particle multiplicity can generate  a "ridge", as well as the evidence for the quark-gluon phase when the emitted particles interact with the medium. For this we use the dual Yang-Mills (YM) theory (see, e.g., [6]) at finite temperature accompanied by the evolution stochastic model of two-particle correlations [7].

{\it Color-electric flux.-} It is known that the symmetry in (thermal) quantum chromodynamics (QCD) vacuum is not as simple one. For example, in the dual Ginzburg-Landau theory (see, e.g., [8,9] and the references therein), the QCD is reduced to $[U(1)]^{2}$ gauge theory including color-magnetic scalar fields, where the latter group originates from the maximal torus subgroup of $SU(3)$.

 At large distances the effective degrees of freedom are light hadrons, e.g., pions, and the corresponding effective theory is based on spontaneously broken chiral invariance. On the other hand, another symmetry present in the chiral limit of QCD is the conformal (or scale) invariance which is broken explicitly by the scale anomaly.

 At the LHC, two protons pass each other in such a way  they can exchange some color messengers, that leads to the connection between them by color strings embedded in the flux tube of color-electric fields. The messenger can be the "dual gluon"- the quanta of the magnetic gauge theory. Here, the confined phase of SU(3) YM theory is described by an effective theory coupling magnetic SU(3) dual gauge potentials $C_{\mu}$ to three adjoint representation scalar fields (see the details in [6]).
The generating current confining the color-electric flux is due a scalar field  (e.g., the dilaton field) squeezing in the  field $C_{\mu}(x)$, and it looks like [10]  $J^{dil}_{\mu} (x)\sim \partial^{\nu}\,G_{\mu\nu} (x)$, where $G_{\mu\nu} = \partial_{\mu}\,C_{\nu} - \partial_{\nu}\,C_{\mu} + \tilde G_{\mu\nu}$, and $\tilde G_{\mu\nu}$ is the Dirac-like string tensor.
The coupling of $C_{\mu}$ to the magnetically charged scalar fields generate color magnetic currents which confine electric flux to narrow tubes connecting a charge and anticharge.

In the paper we follow the scheme [11] where the electroweak (EW) symmetry breaking  at the scale $v$ = 246 GeV (the vacuum expectation value of the $SU(2)$ doublet Higgs scalar field) is triggered by a spontaneous breaking of scale symmetry at an  energy  scale $f\geq v$.
In this scenario, one has new  nearly conformal dynamics at a scale $ \sim 4\pi f$ which can feed into an electroweak sector. The spectrum may contain an EW singlet scalar field $\chi (x)$, the dilaton mode, that is the pseudo-Goldstone boson of spontaneously broken approximate scale invariance.
The dilaton becomes massless when the conformal invariance is recovered. In the extreme mixing scenario $f = v$, the dilaton itself is the Higgs-boson.

The dilaton fields $\chi (x) =\phi (x)\,e^{i\,\omega (x)}$ are associated with not individual particles but the subsidiary objects in massive dual gauge theory. The solution for $C_{\mu}(x)$ 
in terms of the dual gauge coupling $g$ up to the divergence of local phase of the
dilaton field, $\partial_{\mu}\omega (x)$  looks like [10]
$$g\,C_{\mu}(x) = \frac{J^{dil}_{\mu} (x)}{4\,g\,[\phi (x)]^{2}} + \partial_{\mu}\omega (x). $$
At low temperatures $\beta^{-1} =T$ the (finite) energy $F (R,\beta)$ of the isolating string-like flux tube of the length $R$ keeps growing as $R$ (the string tension is unbreakable)
$$F(R,\beta)\sim m^{2}(\beta)\,R [a + b\,\ln(\tilde\mu\,R)], $$
where $m(\beta)$ is the mass of the $C_{\mu}$-field, $\tilde\mu$ is the infra-red mass parameter, $a > 0$ and $b < 0$.

The flux is given in the form
$$\Phi\sim \int C_{\mu}(x)\,dx^{\mu}\sim \frac{1}{g}\,\int\partial_{\mu}\omega (x)\,dx^{\mu}$$
which leads to the requirement that $\omega (x)$ can be varied by $ 2\,\pi\,n$, and the integer $n$ is regarded as the winding number of the flux tube corresponding to the topological charge.
Hence, the flux is quantized as a result of the condition $\Phi\sim 2\,\pi\,n/g$.
Within the differential form of the flux quantization condition [9]
 $$\vec\nabla\times\vec\nabla\omega = 2\,\pi\,n\,\delta (x)\,\delta (y)\,\vec e_{z},$$
where $\omega = n\,\varphi$ and $\varphi$ is the azimuth around the $z$-axis,
the phase of the dilaton field $\omega (x)$ becomes rather singular at the center of the flux.

The  effective theory is applicable at distances greater than the flux tube radius $r\sim m^{-1}(\beta)$. The dilaton fields vanish at the center of the flux tube and approach their vacuum value $f$ (the scale of conformal symmetry breaking) at large distances.  The profile of color-electric field $E$ inside the flux is defined by the rotation of the dual field
$$\vec {E} = \vec {\nabla}\times \vec {C} = \frac {1}{r}\,\frac{d\tilde C (r,\beta)}{dr}\,\vec e_{z} \equiv E_{z}(r,\beta)\, \vec e_{z}, $$
where $\vec e_{z}$ is the unit vector along the $z$-axis, and $E_{z}(r,\beta)$ in terms of $r$-radial coordinate (the distance from the center of the flux) is [10]:
 $$E_{z}(r,\beta) = \sqrt {\frac{\pi\,m(\beta)}{2\kappa\,r}}\,e^{-\kappa\,m(\beta)\,r}\,
 \left [\kappa\,m(\beta) - \frac{1}{2\,r}\right ]$$
with $\kappa\sim O(1)$.
The color-electric strings may be longitudinally stretched and at high temperatures they can break into the separate short pieces as $m(\beta\rightarrow \beta_{c})\rightarrow 0$, where $\beta^{-1}_{c}$ is the critical temperature $T_c \sim O(200~ MeV)$. These short pieces (clusters) then can decay into observable particles, e.g., pions which are finally detected.

The flux solution for $\tilde C(r,\beta)$ - field along the $z$-axis is finite at $r\rightarrow\infty$, $\phi (r) = f$, and has the following transverse behavior [10]:
 $$\tilde C(r,\beta) \simeq \frac{4\,n}{7\,g(\beta)} - \sqrt {\frac{\pi\,m(\beta)\,r}{2\kappa}}\,e^{-\kappa\,m(\beta)\,r}\,
 \left [1 + \frac{3}{8\,\kappa\,m(\beta)\,r}\right ]. $$

Note, that at finite temperature
$$m^{2}(\beta)\sim g^{2}(\beta)\,\delta^{2}(0), $$
where $\delta^{2}(0)$ is the inverse cross section of the flux tube. This cross section becomes large if $m\rightarrow 0$. Hence, at high temperatures, the strings may be stretched also in radial (transverse) direction with $r$. The increasing of $T$ can lead to spreading out of the color-electric flux where the lower bound on $r$ is  $[2\,\kappa\, m(\beta)]^{-1}$. Thus, the strings can expand with $r$ and may break into separate short pieces which then decay into observable particles. It turns out that string pieces move relativistically relative to each other and cannot exchange any messenger.
We suppose that many  flux tubes are chaotically occurred with different transverse size $r\sim m^{-1}(\beta)$, and they are distributed by an arbitrary manner inside the region surrounded by the dilaton fields.

Keeping in mind that the color-electric strings are mainly longitudinal objects the transverse movement of string pieces is due to transversely distorted dual gauge field. The transverse momenta of most of the strings are rather small.
At high temperatures there will be a sufficient amount of transverse flow to provide the collimation around $\Delta\varphi = \varphi_{1} -   \varphi_{2}\simeq 0$. This effect becomes weaker at low $T$ as seen in the CMS at small $\sqrt s$.


{\it Angular correlations.-}   The correlation function $R(\Delta\varphi, \Delta\eta)$ defined in [1] in terms of azimuthal $\Delta\varphi$ and pseudorapidity $\Delta\eta$ separations between two particles is
\begin{equation}
\label{e8}
R(\Delta\varphi, \Delta\eta) = {\langle (\langle N \rangle -1)\, \Lambda (\Delta\varphi, \Delta\eta)\rangle}_{N},
\end{equation}
where $\Lambda $ is the signal faced to two-particle correlation normalized to two one-particle distributions without correlations;  $\Delta\varphi = \varphi_{1} - \varphi_{2}$, $\Delta\eta = \eta_{1} - \eta_{2}$, where $\eta_{i} = - \ln [\tan (\theta_{i}/2)]$, and $\theta_{i}$ being the polar angle of one of the particles ($i=1,2$).  The function $\Lambda$ is
\begin{equation}
\label{e9}
 \Lambda = \lambda_{1}(\beta)\,e^{-\Delta_{k\Re}}
\left [1+\lambda_{2}(\beta)\,e^{ +\Delta_{k\Re}/2}\right ] ,
\end{equation}
where $\lambda_{1} (\beta)\simeq \gamma(\omega,\beta)/(1+\alpha)^{2}$, and
$\lambda_{2}(\beta) \simeq 2\,\alpha/\sqrt{\gamma(\omega,\beta)}$;
 $$\gamma (\omega,\beta) \equiv \gamma (n)  = \frac{{n^2 (\bar \omega )}}{{n(\omega )\ n(\omega
 ')}} ,\ \
 n(\omega ) \equiv  n(\omega ,\beta ) =
 \frac{1}{{e^{\omega \beta} - 1 }} ,\ \
 \bar\omega  = \frac{{\omega  + \omega '}}{2}. $$
 Here, the function $\alpha = \alpha (\beta)$ entering $\lambda_{1}$ and $\lambda_{2}$, the measure of chaoticity, summarizes our knowledge of other than space-time characteristics of the particle emitting source, and it varies from $0$ to $\infty$; $n(\omega,\beta )$ is the mean value of quantum numbers for particles with the energy $\omega$ in the thermal bath with statistical equilibrium at the temperature $\beta^{-1}$.
The shape of the "ridge" depends on the structure of $\Delta_{k\Re}$ - space-time distribution in (\ref{e9}), while its amplitude  - on the  chaotic functions $\lambda_{1}$ and $\lambda_{2}$  which are strongly dependent on $\sqrt {s}$, $\beta$ and $N$.

The distribution of pions can be either far from isotropic, usually concentrated in some directions, or almost isotropic, and what is important that in both cases the particles are under the random chaotic distortion caused by other fields in the thermal medium. In formula (\ref{e9}) all of these features are embedded in both $\lambda_{1}$ and $\lambda_{2}$. Actually, $\Delta_{k\Re} =  (k_{1} - k_{2})^{\mu}\,\Re_{\mu\nu}\,(k_{1} - k_{2})^{\nu}$  is the smearing smooth dimensionless generalized function, where $\Re_{\mu\nu}$ means the nonlocal structure tensor of the space-time size, and it defines the region of emitted particles with four-momenta $k_{1}$ and $k_{2}$.


We have already emphasized [7] that there are two different scale parameters in two-particle correlation physics.
One of them is the correlation radius which gives the "pure" size of the particle emission
source with no  distortion and interaction forces coming from other fields. The other  parameter is the stochastic scale  of the production particle region where the stochastic, chaotic distortion due to environment is enforced.
For practical using  one can replace $\Delta_{k\Re}\rightarrow q^{2}\,L^{2}_{st}$ with $q^{\mu} = (k_{1} - k_{2})^{\mu}$, where
$L_{st} = L_{st}(\beta, M)$ is the stochastic measure of the space-time overlap between two particles with the mass $M$, and the physical meaning of $L_{st}$ depends on the fitting of
$ R(\Delta\varphi, \Delta\eta)  $-function (\ref {e8}).
Taking into account the cylindrical symmetry of the flux around the collision axis and the longitudinal boost invariance, $\Lambda (\Delta\varphi,\Delta\eta)$ (\ref{e9}) becomes
\begin{equation}
\label{e12}
\Lambda = e^{-q^{2}_{T} L^{2}_{T}}\,f_{1} (\alpha, q^{2}_{l}, q^{2}_{0}) +
e^{-q^{2}_{T} L^{2}_{T}/2}\,f_{2} (\alpha, q^{2}_{l}, q^{2}_{0}),
\end{equation}
where
$$ f_{1} = \frac{\gamma (\omega, \beta)}{(1+\alpha)^{2}}e^{-\Delta_{0l}}, \,\,\,
f_{2} = \frac{2\,\alpha\sqrt{\gamma (\omega, \beta)}}{(1+\alpha)^{2}}e^{-\Delta_{0l}/2},$$
$ \Delta _{0l} = L^{2}_{l}[q^{2}_{l} - (q^{0})^{2}] +  [L^{2}_{0} + L^{2}_{l}](q\cdot u)$; $L_{0}$,  $L_{l}$ and $L_{T}$ are the time-like, longitudinal and transverse components of the space-time size $\sqrt {L^{2}_{st}}$; $ q_{T} = \sqrt {(k_{{x}_{1}} - k_{{x}_{2}})^{2} + (k_{{y}_{1}} - k_{{y}_{2}})^{2}}$ and we put $K^{\mu} = (K^{0}, K_{T}, 0, K_{l})$, where
$K^{\mu} =  (k_{1} + k_{2})^{\mu}/2$. Actually, $u_{\mu} (K^{\mu}) = \delta (K^{\mu}) [1,0,0,v_{l} (K^{\mu})]$ is the four-velocity, where $ \delta (K^{\mu}) = 1/\sqrt {1- v^{2}_{l} (K^{\mu})}$ and $v_{l} (K^{\mu})$ is the longitudinal source velocity  (see [12] and references [11,13] therein).
The  following condition $(q\cdot K)= 0$ is evident. The parameters $L_{T},\,L_{l},\,L_{0}$ are invariant under longitudinal boosts and they can be  considered to measure the pion source size of the freeze-out hypersurface. Specifically, $L_{0}$ measures the particle emission duration (lifetime) on which, according to some expectations, the effect of the first order phase transition may be observed.

{\it Source structure.-} In the case of two  emitted pions the transverse size of the source  is ($n\,\beta\,M > 1, n=1,2 ...$)
\begin{equation}
\label{e13}
L_{T} (\beta)\simeq  {\left [\frac {(2\,\pi\,\beta)^{3/2}\,e^{\beta\,M}}
{ 3\,\alpha (N)\,k^{2}_{T}  {\left\vert \tilde\omega /k_{T} - \sqrt {1+  M^{2}/k^{2}_{T}}\right\vert} ^{2}\,M^{3/2}\, \left (1+\frac {15}{8\,\beta\,M}\right )}\right ]}^{{1}/{5}},
\end{equation}
where $k_{T} =\vert\vec k_{T_{1}}+\vec k_{T_{2}}\vert/2$, $\tilde\omega = (\omega_{1} + \omega_{2})/2$, $k^{\mu}_{i}=(\omega_{i},\vec{k}_{T_{i}})$, $i=1,2$.

The  scale $L_{T}$ is varied drastically from its maximal magnitude at small (but finite) $T$ down to its minimum value at the temperature $T$ close to $T_{c}$ above which one can expect a small amount of hadronic states. The analytic criterion constraining the existence of the quark-antiquark bound states at $T > T_{c}$ is obtained in [10], where the Coulomb string tension for strongly interacting particles in the deconfinement phase increases with the strong coupling constant $\alpha_{s}^{2}T^{2}$, and at short distances the Coulomb potential has the linear rising. At high enough $T > T_{c}$ the scale $L_{T}$ increases sharply (explosion) and no hadronic states are already found.

The  squeezing of $L_{T} $ (\ref{e13}) with $T$ leads to decreasing of particles production rate probability. This means the suppression of high $k_{T}$ pions at high multiplicity which is the strong manifestation of the quark-gluon phase stage. It is supported by the interaction of outgoing pions with the medium in thermal bath ($\alpha (N,\beta)$ - dependence).

The "ridge" may appear as the smooth increasing of $\Lambda$-function in ({\ref {e8}).
Because of the long flux-tube with the small transverse size $L_{T}$ the main contribution to the "ridge" is due to (\ref{e12}) with $q^{2}_{T}\rightarrow 0$ ($\Delta\varphi\rightarrow 0$). The increasing of $T$, $k_{T}$ and $N$ events leads to the amplification of the "ridge" effect.

 {\it Conclusion.-} We have studied on the qualitative level the "ridge" pattern in two-particle angular correlations. The effective model based on the dual description of YM theory near the phase transition accompanied by the stochastic model of two-particle correlations has been used. The main results are:

1. The color-electric flux occurred in $pp$ collisions at high $\sqrt {s}$ may break into the separate pieces (clusters) of strings in high thermal medium close to the phase transition. The short color-electric strings decay into pions in radial direction $r\sim m^{-1}(\beta)$.

2. The string pieces move  primarily in high temperature/density matter as a whole in the same direction as two secondary pions emitted out of these short strings with the close azimuthal angles, $\Delta\varphi \simeq 0$.

3. The two pions source is embedded in the thermal bath and this source is disturbed by external forces (in-medium) which are given by the measure of $\alpha$-chaoticity.

4. The angular correlation function $\Lambda (\Delta\varphi, \Delta\eta )$ has no separate rapidity direction along the flux (long-range rapidity availability) and it exhibits an enhancement ("ridge" behavior) only at high $T$ with large amount of multiplicity $N$ events.

5. We can suppose the  quark-gluon phase formation within the interaction of outgoing high $k_{T}$ pions with the medium.

6. This could probe both the size and the temperature of the  source where two pions are emitted into a narrow range of azimuthal angles.




\begin{thebibliography}{30}

\bibitem{1}


 The CMS Collaboration, J. High Energy Phys. 09 (2010) 091.
\bibitem{2}
 E. Shuryak, arXiv: 1009.4635 [hep-ph];  A. Dumitru, et al.,
 Phys. Lett. B697 (2011) 21;  K. Werner, Iu. Karpenko, T. Pierog, arXiv: 1011.0375 [hep-ph].
\bibitem{3}
 B.I. Abelev et al., [STAR Collaboration], Phys. Rev. C80 (2009) 064912; B. Alver et al., [PHOBOS  collaboration], Phys. Rev. Lett. 104 (2010) 142301; A. Adare et al., [PHENIX Collaboration]  Phys. Rev. Lett. 104 (2010) 252301.
\bibitem{4}
M.Yu. Azarkin, I.M. Dremin, A.V. Leonidov, arXiv: 1102.3258 [hep-ph].
\bibitem{5}
B. Andersson, G. Gustaffson, G. Ingelman, T. Sjostrand, Phys. Rep. 97 (1983) 31.
\bibitem{6}
M. Baker J.S. Ball and F. Zachariasen, Phys. Rev. D44 (1991) 3328.
\bibitem{7}
G.A. Kozlov, Phys. Nucl. Part. Lett. 6 (2009) 162.
\bibitem{8}
 H. Suganuma, S. Sasaki, H. Toki, Nucl. Phys. B435 (1995) 207.
\bibitem{9}
 Y. Koma, H. Suganuma, and H. Toki, Phys. Rev. D60 (1999) 074024.
\bibitem{10}
G.A. Kozlov, Phys. Nucl. Part. Lett. 5 (2008) 851.
\bibitem{11}
E. Gildener and S. Weinberg, Phys. Rev. D13 (1976) 3333.
\bibitem{12}
 K. Morita, S. Muroya, H. Nakamura, and C. Nonaka, Phys. Rev. C61 (2000) 034904.
\end{thebibliography}
\end{document}